\documentclass[aps,prl,amsmath,amssymb,preprint,superscriptaddress]{revtex4-2}

\usepackage{graphicx}
\graphicspath{{figures/}}
\usepackage{color}
\usepackage{float}
\usepackage{soul}
\usepackage{dcolumn}
\usepackage{bm}
\usepackage{xcolor} 
\usepackage[unicode,colorlinks=true]{hyperref}
\usepackage{upgreek}
\usepackage{siunitx}
\usepackage[version=4]{mhchem}  
\setcitestyle{super}

\begin{document}

\section{Title}
Ultrafast tuning of the focusing efficiency of a nonlinear atomically thin lens

\section{Author list}
Rahil Rezwan$^{\star,1,2}$, Bernardo Dias$^{\star,3}$, Tom Hoekstra$^{3}$, Mehmet Atıf Durmuş$^{3}$, Bauke van der Vorm$^{3}$, Till Weickhardt$^{1}$, Omid Ghaebi$^{1}$, Zhuoyuan Lu$^{4}$, Devapriyo Mithun$^{1}$, Carsten Ronning$^{1,5}$, Yuerui Lu$^{4}$, Jorik van de Groep$^{3}$ and {Giancarlo} Soavi$^{1,5,\dagger}$

\section{Affiliations}
\noindent
$^1$Institute of Solid State Physics, Friedrich Schiller University Jena, 07743 Jena, Germany
\newline
$^2$ARC Centre of Excellence for Transformative Meta-Optical Systems, Department of Electronic Materials Engineering, Research School of Physics, The Australian National University, Canberra, ACT 2601 Australia
\newline
$^3$Van der Waals-Zeeman Institute, Institute of Physics, University of Amsterdam, Amsterdam 1098 XH, The Netherlands
\newline
$^4$School of Engineering, College of Engineering, Computing and Cybernetics, Australian National University, Canberra ACT 2601, Australia
\newline
$^5$Abbe Center of Photonics, Friedrich Schiller University Jena, 07745 Jena, Germany
\newline
$^{\star}$ These authors contributed equally.
\newline
$^{\dagger}$ email: giancarlo.soavi@uni-jena.de

\begin{abstract}
Control over light focusing is paramount for optical and imaging systems. However, its implementation in active and integrated nanophotonic devices is hindered by the bulky and static nature of standard lenses. Here, we show that Fresnel Zone Plate Lenses (FZPL) based on a monolayer Transitional Metal Dichalcogenide are a viable solution to the challenges of miniaturization, integration, and active control of the focusing efficiency, featuring both ultrafast and large modulation depth. Using a monolayer $\ce{WSe2}$ FZPL, we demonstrate all-optical ultrafast ($\sim$ ps)  modulation of the focusing efficiency close to $\sim$30\%, that we achieve by exploiting the nonlinear exciton-resonant enhancement and modulation of second harmonic generation. 
\end{abstract}

\maketitle

\section{Introduction}

The ability to control and focus light is paramount to a wide range of optical and imaging systems, including microscopes, telescopes, and cameras. To realize the spatial phase profile required for focusing, bulk optical lenses rely on their physical size and curvature, limiting their integration into miniaturized and lightweight devices.~\cite{lohmann1989scaling,yu2014flat}. Metasurface flat optical elements and Fresnel zone plate lenses (FZPL) provide powerful alternatives for nanophotonic and on-chip compatible technologies~\cite{suwannasopon2019miniaturized, brongersma2020road, pendry2000negative, arbabi2015subwavelength}. Specifically, a FZPL is a diffractive optical element that is comprised of alternating transparent and opaque concentric rings designed to focus light through constructive interference of the diffracted wave fronts originating from the micro-scale apertures. This design is well-established in EUV and X-ray optics, as conventional refractive optics are hindered by the high absorption and near-unity refractive index of materials in this spectral regime~\cite{baez1961fresnel, overbuschmann2012fabrication, bach1998properties}. While, FZPLs based on absorptive metals are commonly used in EUV and soft X-ray microscopy~\cite{ossiander2023extreme, sun2022procedure, wu2012hard}, photomask metrology~\cite{perera2022development}, and infrared detection~\cite{rossi2000diffractive}, their thickness still ranges from hundreds of \SI{}{\nano\metre} to a few \SI{}{\micro\metre}~\cite{wu2012hard} in order to achieve sufficient optical contrast. Recently, flat lenses with a thickness reaching down to the atomic limit have been demonstrated in the visible and near-infrared range, capitalizing on the strong light-matter interactions offered by 2D semiconductors~\cite{azimi2025photonics, de2025roadmap, qin2021pi}. In contrast to established metalenses based on plasmonic~\cite{maier2007plasmonics} or Mie resonances~\cite{kuznetsov2016optically}, these atomically thin flat lenses (ATFL) employ the conventional FZPL design to concentrate incident light into a focal point with the additional benefit that they can be transferred on any substrate~\cite{Yang2016AtomicallyThin, liu2018ultrathin}. Uniquely, the focusing efficiency of ATFL can be actively manipulated by tuning the excitonic resonances in 2D transition metal dichalcogenides (TMDs)~\cite{van2020exciton, park2021focus}. This can be rationalized by considering that the focusing efficiency is directly proportional to the refractive index of the concentric rings~\cite{anivcin1989fresnel, rastani1991binary, tatchyn1985symmetry}, and modulation of the optical properties of the ring material directly tunes the diffraction efficiency and thereby the focal intensity. So far, dynamical modulation in an ATFL has been achieved by electrical~\cite{van2020exciton} and thermal~\cite{guarneri2024temperature} means, achieving a change in the focusing efficiency of $>30\%$. Nevertheless, these devices are crucially limited by their low modulation speeds on the order of ms~\cite{van2020exciton}.

All-optical modulation offers an attractive alternative to thermal and electrical tuning for its inherent ultrafast modulation speeds. In this approach, light transmission is modulated by a control beam (CB) that bleaches an optical resonance, with the switching speed limited only by the excited state lifetime. In atomically thin TMDs, the typical exciton recombination dynamics range from hundreds of ps down to sub-ps lifetimes~\cite{pogna2016photo, sim2013exciton}. Yet, this comes at the cost of a small modulation depth, with typical reported values of the normalized differential transmission $\Delta T/T <1\%$, even close to the saturating fluence of $>$\SI{50}{\micro\joule\per\centi\metre\squared}~\cite{seo2016ultrafast, li2021revealing}. This highlights the inherent trade-off between high modulation depth but slow modulation speed achieved by electrical and thermal tuning~\cite{newaz2013electrical, yu2017giant, liu2020temperature}, versus low modulation depth but ultrafast modulation achieved with all-optical methods~\cite{pogna2016photo, seo2016ultrafast}. 

In this context, higher-harmonic generation is a powerful method to achieve the best of both worlds: strong and rapid modulation at the same time~\cite{klimmer2021all}. Due to their broken space inversion symmetry, monolayer TMDs have a strong second-order nonlinear susceptibility~\cite{dogadov2022parametric}, which is highly enhanced at exciton resonances~\cite{wang2015giant}. By harnessing the exciton's tunability, strong and ultrafast all-optical modulation of the second-harmonic intensity can be achieved, with $>50 \%$ modulation and sub-ps speeds already demonstrated~\cite{taghinejad2020photocarrier}. Here, we leverage this effect for ultrafast all-optical tuning of the focusing efficiency of a monolayer $\ce{WSe2}$ flat lens in the nonlinear regime of second-harmonic generation (SHG) (Fig. \ref{fig:fig1}(a)), achieving both large modulation of the focusing efficiency (close to 30\% at a fluence of $<$\SI{10}{\micro\joule\per\centi\metre\squared}) and ultrafast sub-\SI{100}{\pico\second} modulation speed. Our results pave the way towards more complex miniaturized photonic devices featuring atomically thin flat lenses as active and tunable optical elements.

\section{Results and Discussion}

\begin{figure*}[h]
\centering
\includegraphics[width=\linewidth]{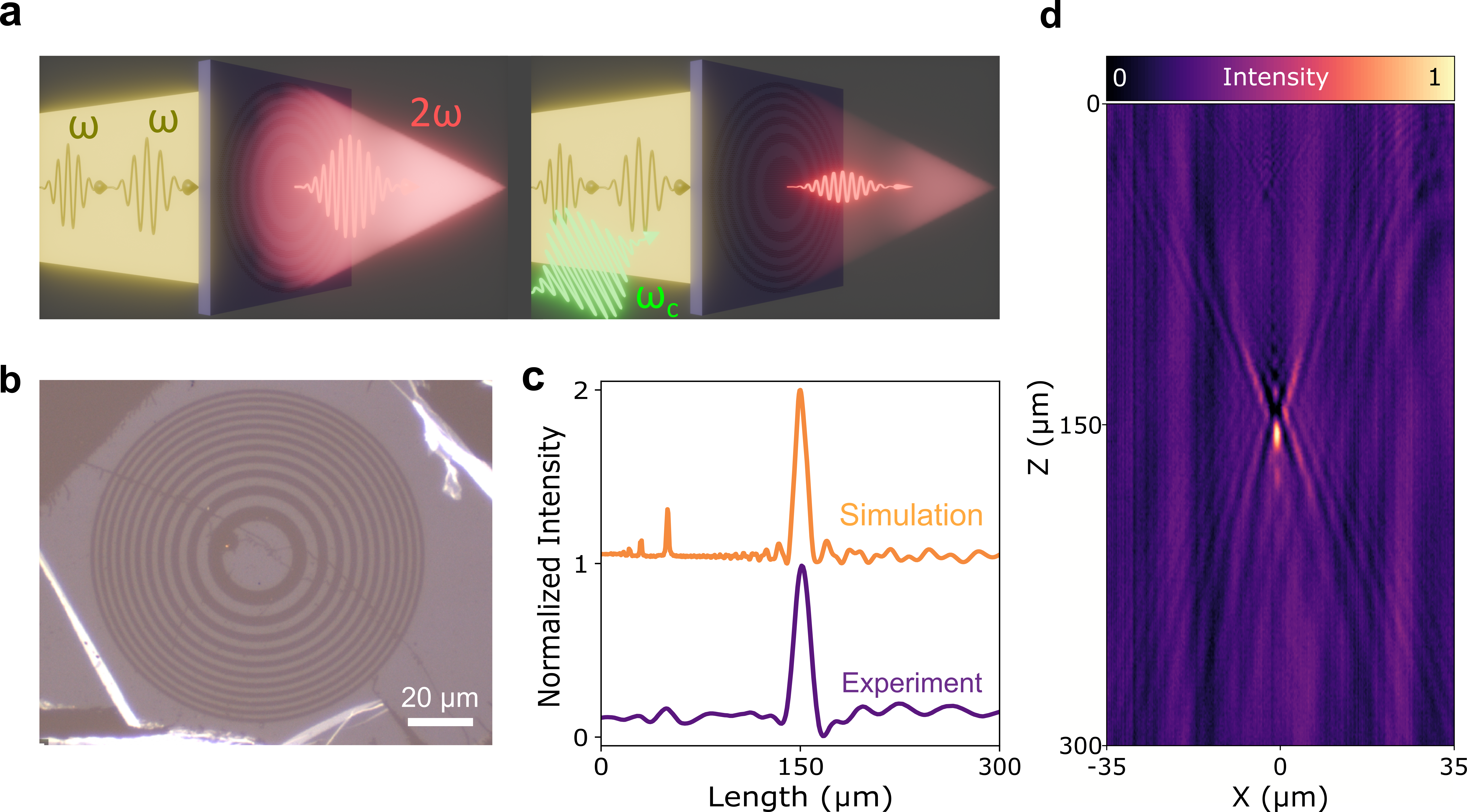}
\caption{\fontsize{10pt}{5pt}\selectfont\label{fig:fig1}
a) Schematic representation and working principle of the all-optical modulation of the SH focusing efficiency by the ATFL. b) Optical image of the fabricated ATFL. The darker gray rings correspond to the the etched areas where the underlying quartz substrate is exposed, while the brighter regions correspond to the monolayer $\ce{WSe2}$. c) Normalized numerical and experimental intensity along the optical axis highlighting the focal length and the profile of the fabricated lens when operated in the linear regime. For the ease of visualization, the value of the simulated result is vertically shifted by 1. The peaks below \SI{150}{\micro\metre} are due to higher diffraction orders. d) Intensity profile in the x-z plane when a collimated beam at \SI{750}{\nano\metre} is incident on the lens.
}
\end{figure*}

To achieve ultrafast tuning of the focusing efficiency of a FZPL, we design a lens composed of monolayer $\ce{WSe2}$ to focus the SHG radiation (Fig.\ref{fig:fig1}a). In this material, the exciton resonance at 755 nm strongly enhances the second-order nonlinear susceptibility, driving efficient SHG at 1510 nm and making the focusing efficiency (at 755 nm, double the frequency) highly sensitive to the exciton population. As such, to control the focusing efficiency, a pump pulse at 515 nm excites carriers across the bandgap, bleaching the exciton ground state and thereby suppressing the resonant enhancement of the SHG on an ultrafast timescale set by the exciton recombination dynamics.

The monolayer $\ce{WSe2}$ is obtained by Au-assisted exfoliation onto a 1 mm thick quartz substrate\cite{AuExfol}. The FZPL is then patterned by electron beam lithography, yielding a lens with a diameter of \SI{100}{\micro\metre} consisting of 22 concentric rings (Fig. \ref{fig:fig1}b). We design the lens to operate at 755 nm with a focal length of \SI{150}{\micro\metre}, matching the lowest exciton resonance of monolayer $\ce{WSe2}$, with zone radii following the binary FZPL formulation $r_n \approx \sqrt{n\lambda_{opt}f}$, where $\lambda_{opt}$ and $f$ are the operating wavelength and focal length, respectively~\cite{young1972zone}. To benchmark the fabricated lens, we first characterize its focusing properties in the linear regime under collimated illumination at \SI{750}{\nano\metre}. We obtain excellent agreement between the simulated (\SI{150}{\micro\metre}) and experimental (\SI{152}{\micro\metre}) focal length (Fig. \ref{fig:fig1}c,d), and measure a focusing efficiency of $\sim$0.06\%, consistent with values reported for monolayer TMD zone plates~\cite{van2020exciton, guarneri2024temperature}. The small efficiency stems from the \SI{6.49}{\angstrom} thickness of the $\ce{WSe2}$ rings~\cite{wilson1969}, which are therefore largely transparent and only weakly scatter the incident field.

\begin{figure*}[h]
\centering
\includegraphics[width=\linewidth]{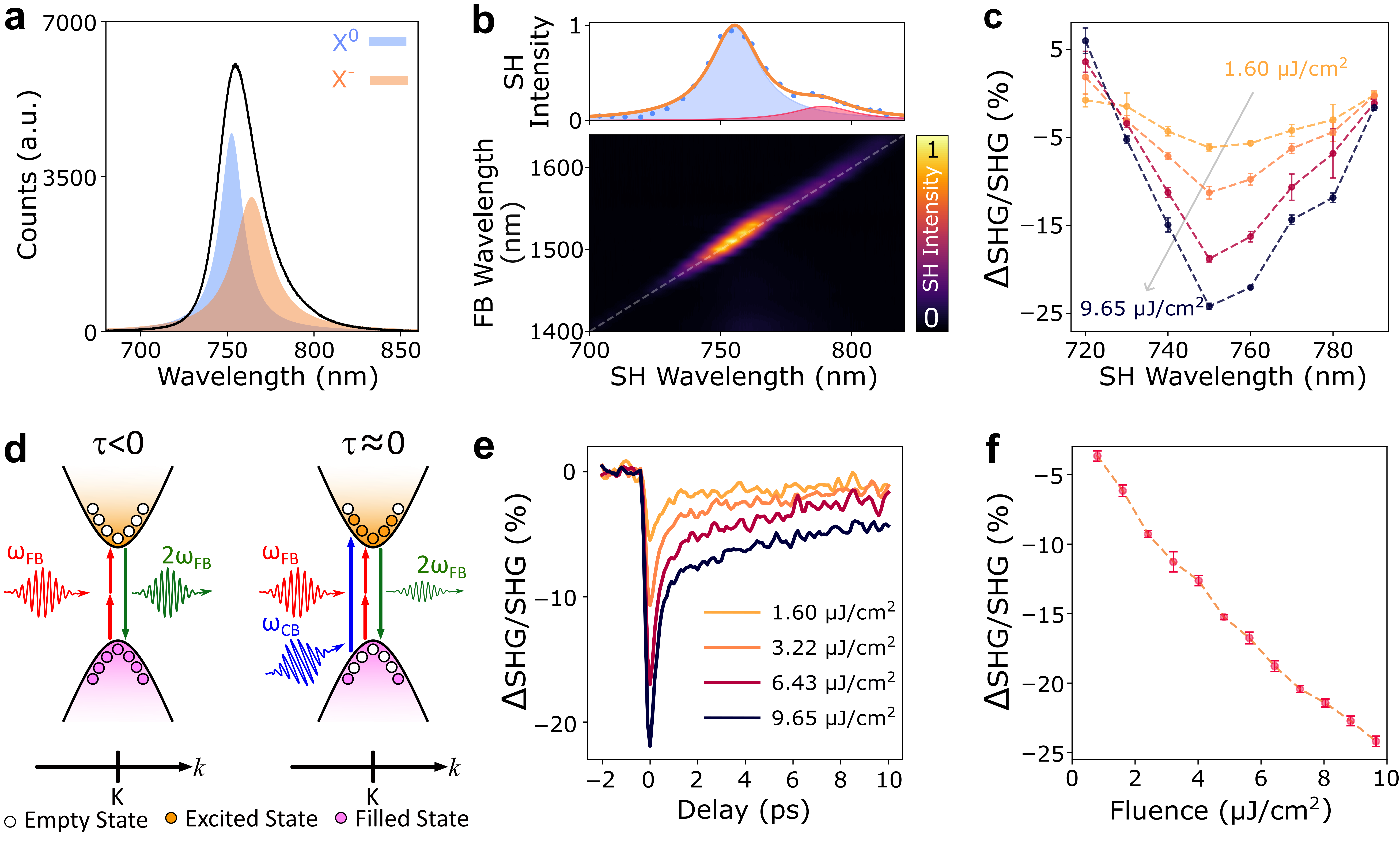}
\caption{\fontsize{10pt}{5pt}\selectfont\label{fig:fig2} 
a) PL spectra of a continuous monolayer $\ce{WSe2}$ in the central area of the ATFL. b) On bottom, fundamental beam (FB) wavelength-dependent SH spectra of monolayer $\ce{WSe2}$. On top, the normalized maximum SH counts, clearly indicating exciton resonant enhancement at $\sim$\SI{755}{\nano\metre}. c) Wavelength dependent maximum SH modulation for different fluence of the CB. d) Schematic illustration of the nonlinear modulation scheme. e) Ultrafast SH modulation of monolayer $\ce{WSe2}$ at \SI{750}{\nano\metre} for different fluences of the CB. f) Fluence-dependent maximum SH modulation depth at \SI{750}{\nano\metre}.
}
\end{figure*}

Since the ATFL operates via SHG intrinsic to the $\ce{WSe2}$ monolayer, we characterize the linear and nonlinear optical response of the bare monolayer in the central region of the lens, avoiding the patterned rings and thus any lensing effects (Fig.\ref{fig:fig2}a,b). Photoluminescence (PL) from this region shows a well-defined peak around 755 nm, confirming the monolayer nature of the $\ce{WSe2}$ layer (Fig.\ref{fig:fig2}a)~\cite{tonndorf2013photoluminescence}. We fit the PL spectrum with two Lorentzian functions, corresponding to neutral exciton (X$^0$) and trion (X$^-$) contributions, with linewidths of 36 and 56 meV, respectively. Next, we assess the second-harmonic response of the bare $\ce{WSe2}$ monolayer by focusing a tunable pulsed laser between \SI{1400}{\nano\metre} and \SI{1640}{\nano\metre}, and observe a strong enhancement of the SH intensity when the generated harmonic overlaps the exciton resonance at $\sim$\SI{755}{\nano\metre} (Fig.~\ref{fig:fig2}b). 
Taken together, the PL and SHG measurements establish how the exciton strongly affects the SHG process in monolayer $\ce{WSe2}$, which forms the basis for the ATFL operation and its ultrafast optical control.

To investigate how an above-bandgap \SI{515}{\nano\metre} CB modulates the exciton-enhanced SH response, we a perform pump–probe study for a large range of fluences from 1.60 to \SI{9.65}{\micro\joule\per\centi\metre\squared} (Fig.~\ref{fig:fig2}c). When illuminating the monolayer with the fundamental beam (FB) at $\omega_\mathrm{FB}$, we observe a strong SHG resonantly enhanced by the excitonic second-order susceptibility. Upon pumping by the CB at $\omega_\mathrm{CB}$, carriers are promoted to the conduction band and bleach the ground-state~\cite{cheng2020ultrafast}, which suppresses the SH intensity (Fig.~\ref{fig:fig2}d), consistent with previous reports of ultrafast exciton bleaching in TMD monolayers~\cite{cheng2020ultrafast, peterka2023high}. The maximum modulation depth for all CB fluences is achieved at $\sim $\SI{750}{\nano\metre}, in correspondence with the exciton resonance wavelength. At the same time, blueshifting of the excitonic resonance due to phase-space filling and electronic screening~\cite{WS2MottDiscontinuous}, combined with a broadening of the Fermi-Dirac distribution due to an increase of the lattice and electronic temperatures~\cite{ghaebi2024ultrafast, soavi2019hot}, lead to an enhancement of the SH intensity at \SI{720}{\nano\metre}. As the photoexcited carriers relax and recombine to the ground state, the SH intensity recovers to its original value (Fig.~\ref{fig:fig2}e). We fit the temporal SHG dynamics with a bi-exponential decay, where the fast component is typically attributed to internal excited state relaxation~\cite{trovatello2020ultrafast}, while the slow component accounts for the total recombination time, including both radiative and non-radiative decay channels~\cite{pollmann2015resonant, steinleitner2017direct, cui2014transient, moody2016exciton}. From the fitting, we obtain decay times of $<$\SI{0.5}{\pico\second} for the fast component and $\approx$ \SI{10}{\pico\second} for the slow component, in good agreement with values reported in literature~\cite{ceballos2016exciton}.

\begin{figure*}[h]
\centering
\includegraphics[width=\linewidth]{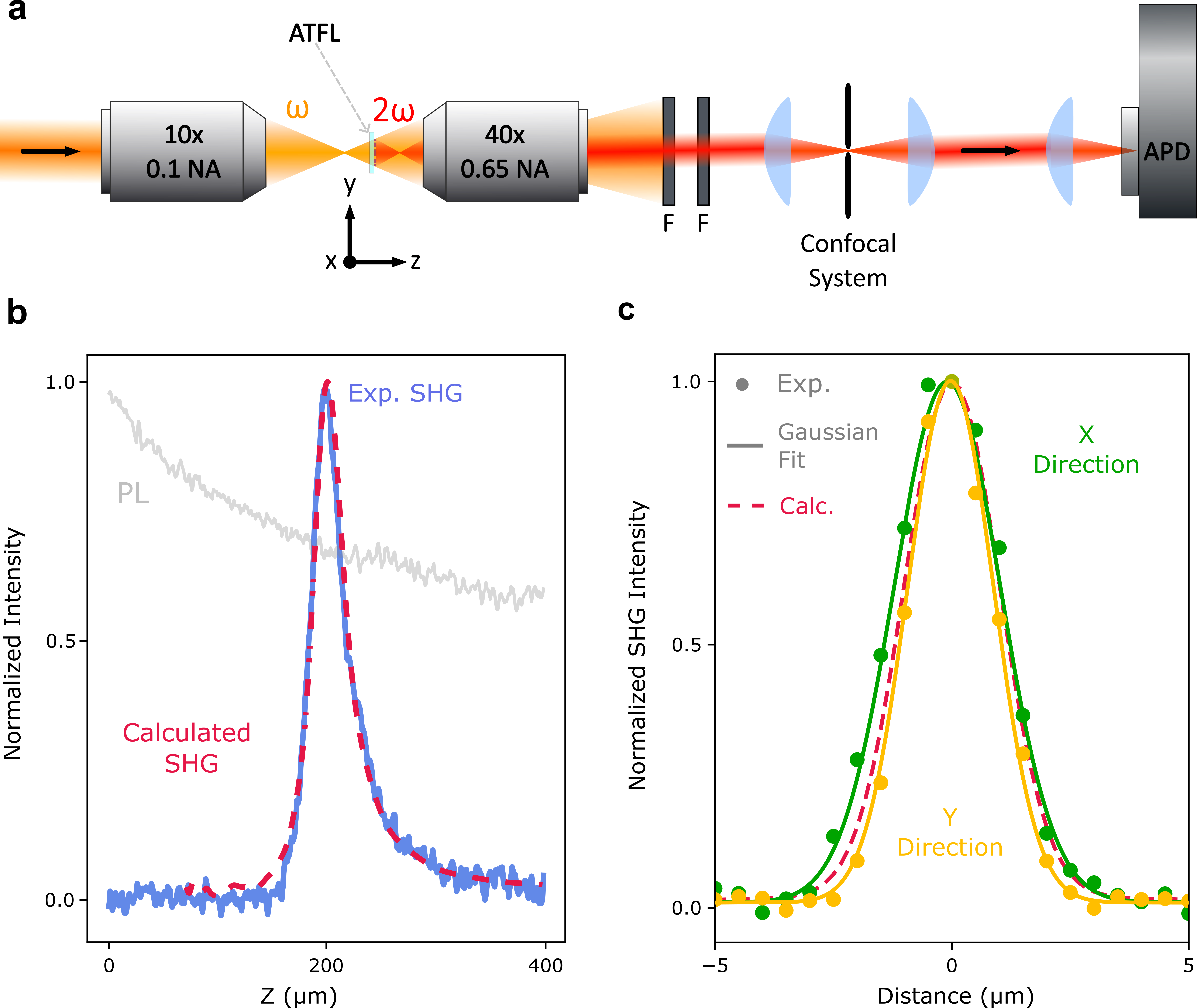}
\caption{\fontsize{10pt}{5pt}\selectfont\label{fig:fig3}
a) Sketch of the setup used for the characterization of the ATFL in the nonlinear regime of SHG. Excitation beam is matched to lens diameter. b) Detected SH (blue curve) as a function of distance from the lens. The deviation from the design focal length (\SI{150}{\micro\metre}) and asymmetrical Gaussian profile is caused by the diverging behavior of the excitation beam. The calculated SH focusing profile using Rayleigh-Sommerfeld diffraction integral is represented by the red dashed curve. The incoherent PL (gray line) is not focused. c)  Characterization of the focal spot shape in the horizontal (X) and vertical (Y) directions. The red dashed curve shows the calculated spot shape.
}
\end{figure*}

Having characterized the nonlinear response of the bare $\ce{WSe2}$ monolayer, we now demonstrate focusing of the generated SH light by the ATFL. Since SHG preserves the spatial and temporal coherence of the fundamental wave~\cite{lochner2019controlling}, the SH signal generated by the odd rings of the lens is focused in the same way as a coherent field at that wavelength. To match the FB beam diameter to the lens aperture (\SI{100}{\micro\metre}), we place the ATFL past the FB focal position, where the beam is diverging (Fig.\ref{fig:fig3}a). The transmitted SH signal is then collected through a confocal pinhole system while scanning the ATFL along the optical axis. The resulting SH intensity profile (Fig.\ref{fig:fig3}b, blue curve) shows clear focusing behavior, with a measured focal length of $\sim$\SI{200}{\micro\metre}. The deviation from the designed focal length of \SI{150}{\micro\metre} and the asymmetric axial profile are both attributed to the diverging profile of the FB, and are well reproduced by diffraction integral calculations taking into account the illumination conditions (Fig.\ref{fig:fig3}b, red dashed line). We note that two-photon PL due to excitation by FB and one-photon PL due to excitation by CB are also generated alongside SHG, but they do not contribute to the focusing due to their incoherent nature~\cite{herrmann2023nonlinear, klimmer2026probing}.

To complete the steady-state characterization, we measure the focal spot size by scanning the ATFL laterally at the experimental focal length position (Fig.\ref{fig:fig3}c). The spot profiles along the horizontal ($x$) and vertical ($y$) directions show FWHM values of \SI{2.73}{\micro\metre} and \SI{2.11}{\micro\metre}, respectively. Here, the slight asymmetry in FWHM is attributed to a small tilt in the illumination and the polarization dependence of SHG\cite{wang2015giant}. The nonlinear focusing efficiency, defined as the ratio of SH power at the focal point to total SH power generated by the ATFL, is $\sim$9.3\% (see Methods for details), more than 150 times higher than in the linear regime. This crucial improvement arises from the nonlinear SH signal that is generated exclusively by the material (odd rings), providing near-ideal amplitude contrast between the even and odd rings.

\begin{figure*}[h]
\centering
\includegraphics[width=\linewidth]{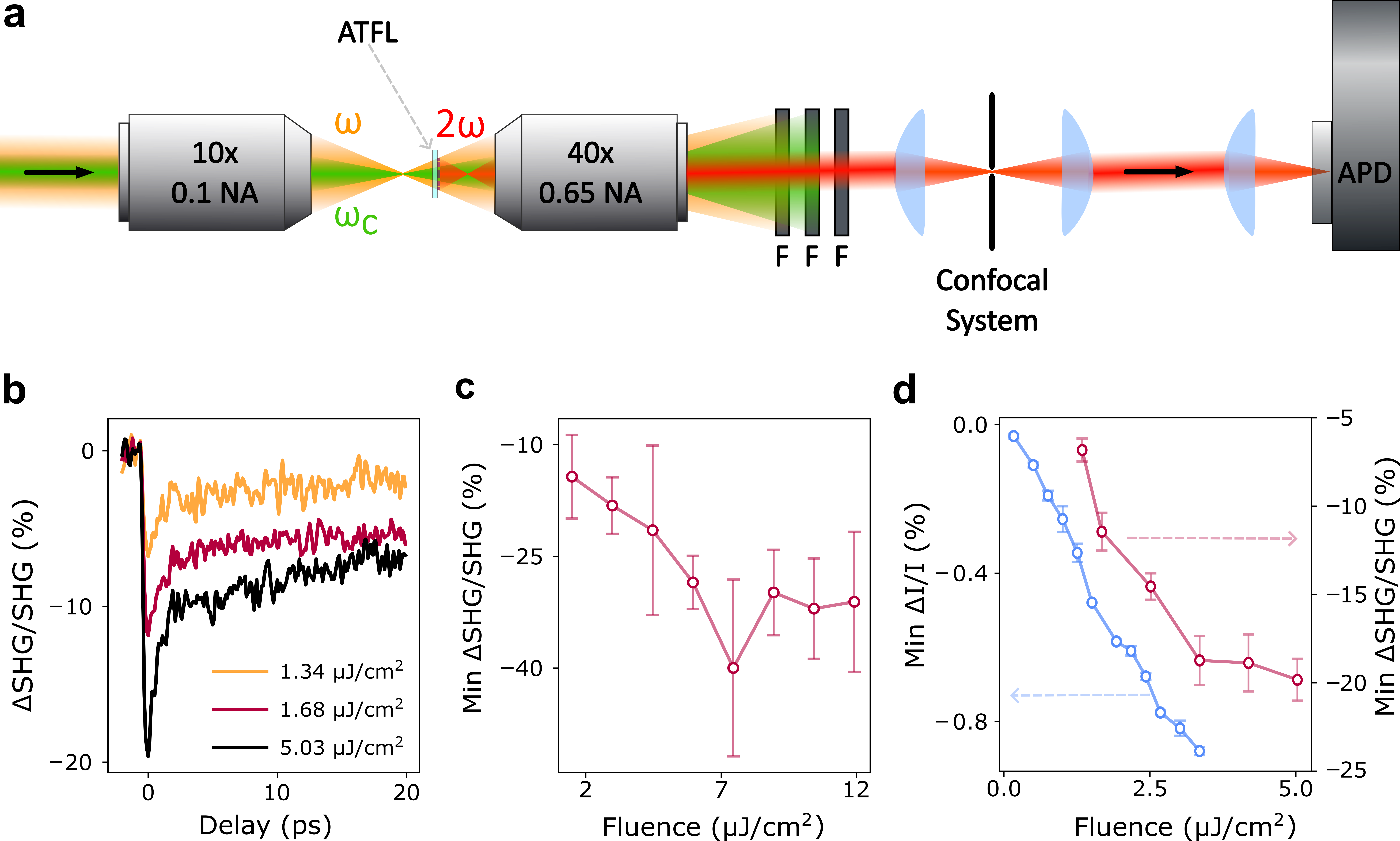}
\caption{\fontsize{10pt}{5pt}\selectfont\label{fig:fig4}
a) Schematic illustration of the setup for the ultrafast all-optical modulation of the ATFL SH focusing efficiency. b) Ultrafast modulation of SH for different fluences. c) Maximum modulation depth as a function of CB fluence. Error bars are calculated as the standard deviation among three measurements. d) Comparison between linear (blue) and second-order (red) modulation depth.
}
\end{figure*}

Having established the steady-state nonlinear focusing, we finally demonstrate ultrafast all-optical modulation of the ATFL focal efficiency in a pump–probe configuration (Fig.\ref{fig:fig4}a). Again, the \SI{515}{\nano\metre} CB induces exciton bleaching across the entire lens aperture, suppressing the SH intensity at the focal point. Figure \ref{fig:fig4}b shows the temporal evolution of the focal SH intensity for different CB fluences, closely mirroring those of the bare monolayer (Fig.\ref{fig:fig2}e), confirming that the same bleaching mechanism governs the lens response. Crucially, operating in the nonlinear regime allows this ultrafast modulation to be combined with a large modulation depth of $\sim$30\% at CB fluences below \SI{10}{\micro\joule\per\centi\metre\squared} (Fig.\ref{fig:fig4}c). This is in stark contrast to the linear regime, where the same CB induces a focal intensity modulation of $<$1\% (Fig.~\ref{fig:fig4}d), highlighting how the resonant enhancement of the nonlinear susceptibility at the exciton transition amplifies the modulation of the focusing efficiency by more than an order of magnitude.

\section{Conclusion}
In conclusion, we have demonstrated ultrafast all-optical modulation of the focusing efficiency of a monolayer $\ce{WSe2}$ ATFL operating in the nonlinear regime of SHG. The lens consists of 22 concentric $\ce{WSe2}$ rings designed to operate at the exciton resonance of \SI{755}{\nano\metre}, where the second-order susceptibility is strongly enhanced. By exploiting this resonant enhancement, an above-bandgap \SI{515}{\nano\metre} CB induces exciton bleaching across the entire lens aperture, directly modulating the nonlinear susceptibility and thus the focal intensity on an ultrafast timescale.

The ATFL focuses the generated SH signal with a measured focal length of $\sim$\SI{200}{\micro\metre}, with a nonlinear focusing efficiency of $\sim$9.3\%, more than 150 times higher than in the linear regime. All-optical modulation of the focal intensity yields a modulation depth of $\sim$30\% at CB fluences below \SI{10}{\micro\joule\per\centi\metre\squared}, with a modulation speed of $\sim$ps. This represents nine orders of magnitude improvement in switching speed over previously reported electrical tuning~\cite{van2020exciton}, while maintaining a comparable modulation depth~---~a combination unachievable in the linear regime, where the same CB induces $<$1\% focal intensity modulation.

These results establish nonlinear ATFLs as a compelling platform for ultrafast active nanophotonics, where the resonant enhancement of nonlinear optical processes provides a powerful handle on both focusing efficiency and modulation depth simultaneously. Straightforward pathways to further improvement include operation in higher-order nonlinear regimes~\cite{cheng2020ultrafast, wang2022optical, peterka2023high} and integration of multiple ATFLs into on-chip photonic architectures, opening prospects for ultrafast beam steering, optical switching, and miniaturized nonlinear imaging systems.

\section{Methods}
\subsection{Au Exfoliation of Monolayer $\ce{WSe2}$}
We exfoliate monolayer $\ce{WSe2}$ using the  previously reported Au-assisted exfoliation procedure~\cite{AuExfol, AuExfol_Science}.
Using electron beam physical vapor deposition (Polyteknik Flextura M508 E), a \SI{100}{\nano\metre} Au layer is deposited on a cleaned Si wafer at a rate of \SI{.5}{\angstrom\per\second}. This procedure is followed by spin coating of a \SI{300}{\nano\metre} layer of PMMA on top of the $\ce{Au}$, which provides mechanical stability for detachment from the $\ce{Si}$ wafer. We use thermal release tape (TRT, Revalpha RAY-4LSC(N), Nitto Denko Corporation) in contact with the PMMA/Au layers to peel from the wafer, which is pressed immediately against a freshly cleaved bulk $\ce{WSe2}$ crystal (HQ Graphene). Separation from the bulk crystal leaves a large-area monolayer on the $\ce{Au}$, which we put in contact with base piranha-cleaned $\ce{SiO2}$ substrates. The substrates are then heated to \SI{110}{\celsius} to remove the TRT, cleaned with acetone to remove the $\ce{PMMA}$ and placed in $\ce{Au}$ etchant (651818, Sigma–Aldrich) for 2 minutes to remove the gold. The sample is washed in isopropanol and dried with a nitrogen gun. 
We observed that after exfoliation the PL is not stable over long measurements, possibly due to defects and residue at the top interface. We thus treated the monolayer with oleic acid at \SI{140}{\celsius} followed by IPA rinse and $\ce{N2}$ dry, which has been reported to passivate defects and stabilize emission\cite{OleicAcid1, OleicAcid2}. 

\subsection{ Sample Fabrication}
Before resist coating, the monolayer sample was cleaned by rinsing in acetone and isopropanol for \SI{60}{\second} each. To form a positive resist layer, the sample was spin-coated with polymethyl methacrylate (PMMA 950K A8, diluted 1:4 in anisole) and baked for \SI{60}{\second} at \SI{180}{\celsius}. To reduce charging during exposure, a conductive polymer layer, Elektra 92 (AR-PC 5092.02), was spin-coated on top and baked for \SI{60}{\second} at \SI{90}{\celsius}. The lens design was patterned in the resist using electron-beam lithography (Raith Voyager), with an acceleration voltage of \SI{50}{\kilo\volt}, a beam current of \SI{1.97}{\nano\ampere}, and a dose of \SI{231.86}{\micro\coulomb\per\centi\metre\squared}. Following exposure, the sample was developed for \SI{15}{\second} in $\ce{H2O}$ and \SI{47}{\second} in MIBK:IPA (1:3), rinsed in isopropanol, and blow-dried with $\ce{N2}$. Using the patterned resist as the etch mask, the monolayer was etched for \SI{28}{\second} by $\ce{O2}$ descum (Oxford PlasmaPro 80) with an RF power of \SI{50}{\watt}. Lift-off was performed by soaking the sample in acetone for \SI{1}{\hour} at \SI{50}{\celsius} and \SI{2}{\hour} at room temperature, followed by rinsing in isopropanol and blow-drying with $\ce{N2}$.

\subsection{PL Characterization}
For PL characterization, the central area of the ATFL containing the monolayer $\ce{WSe2}$ was excited with a \SI{532}{\nano\metre} diode laser (Cobolt 08-DPL \SI{532}{\nano\metre}). The light was focused on the sample with a 50x objective (M Plan Apo 50x, Mitutoyo) and the PL, collected in reflection geometry, was guided to a spectrometer (Horiba iHR550) after filtering by a longpass dichroic mirror (LPD02-532RU-25, Semrock) and a notch filter (FL532-3, Thorlabs).

\subsection{Focal Length Measurement}
To characterize the focal length, we built a setup operating as a confocal microscope. The primary laser source (FLINT12 from Light Conversion) is a mode locked Yb laser (wavelength \SI{1030}{\nano\metre}, repetition rate of \SI{76}{\mega\hertz}, power \SI{12}{\watt}, pulse duration $\sim$\SI{100}{\femto\second}). A portion (\SI{4.5}{\watt}) of this laser pumps an Optical Parametric Oscillator (OPO) (Levante fs IR from APE), which provides a tunable output from \SI{1300}{\nano\metre} to \SI{2000}{\nano\metre} for the signal and from \SI{2150}{\nano\metre} to \SI{4050}{\nano\metre} for the idler, both with pulse duration of $\sim$150 - \SI{200}{\femto\second}, depending on the wavelength. The signal beam of the OPO further pumps a second harmonic generation system (HarmoniXX SHG from APE), which provides output from \SI{650}{\nano\metre} to \SI{1000}{\nano\metre}. The diameter of this SH beam (\SI{750}{\nano\metre}) was reduced to $\approx$\SI{600}{\micro\metre} with a home-built beam reducer. The ATFL was illuminated by a beam at \SI{750}{\nano\metre} and, on the transmission side, the light was collected by a 40x objective with numerical aperture of 0.65 (40X Plan Achromat, Olympus). Additionally, the collected light is focused into a \SI{5}{\micro\metre} pinhole and collected again with a lens (C240TMD and LA1422-B, Thorlabs). This system limits the detected light other than that coming from the focal plan of the collection objective, in analogy with a confocal microscope. Finally, the light is detected with an Avalanche Photodetector (APD130A/M from Thorlabs) after filtering (FBH750-10, Thorlabs). The position of the ATFL can be changed with motorized stages (OSMS80-20ZF-0B and TAMM100-50c(XY) from OptoSigma). By changing the position of ATFL along the optical axis, we recorded the intensity profile of the diffracted light. In case of nonlinear (SHG) focal length measurement, the sample was illuminated directly with the signal beam of the OPO. The intensity profile of the diffracted SHG is recorded in the the same was as described above. Data acquisition scripts for the measurements of the focal length were generated using OpenAI ChatGPT (accessed July 2025). All AI-generated code was reviewed, tested, and revised by the authors, who confirmed accuracy and reproducibility of results.

\subsection{Static and Dynamic Wavelength Dependence}
For the static wavelength dependence, we used the tunable OPO signal in the wavelength range \SI{1400}{\nano\metre} to \SI{1640}{\nano\metre} and a custom-built optical microscope in transmission geometry. The collected light was guided to a spectrometer consisting of a monochromator (Horiba iHR 320) and a liquid nitrogen-cooled silicon (Si) CCD detector (Horiba Symphony II) after filtering (FESH0950 from Thorlabs).
For the time-resolved studies, the sample was pumped by an additional laser beam at a  wavelength of \SI{515}{\nano\metre}, acting as the pump/CB, obtained from the SH of the primary Yb laser source. The pump and FB were combined in a dichroic mirror (DMSP650 from Thorlabs), and their relative delay was controlled by a motorized stage (M-404.2PD from PI). The light was collected in transmission geometry and was measured with an APD (APD440A/M, Thorlabs) after filtering.

\subsection{Numerical Simulations}
The numerical simulations for the focal length measurements were performed in Python using the LightPipes package, a simulation tool for propagation, diffraction and interference of coherent light~\cite{lightpipes}. The simulation was done by initializing a square grid size of 2000x2000, wavelength of \SI{750}{\nano\metre} with Gaussian beam profile, setting a \SI{500}{\nano\metre} resolution and the LightPipes Fresnel propagation algorithm. The ATFL structure was defined with the equation $r_n\approx\sqrt{n\lambda_{opt} f}$ for the n-th zone with a total of 22 zones, resembling our fabricated structure.

\subsection{Calculation of Nonlinear Focusing Efficiency}
For calculating the nonlinear focusing efficiency, we have calculated the sheet SH tensor element $\chi_S^{(2)}$ from the average applied FB and SH power obtained from the center of the ATFL according to the equation~\cite{woodward2017characterization}:
$$\chi_S^{(2)}=\sqrt{\frac{c^3\epsilon_0f\pi r^2 t_{FWHM}(1+n_2)^6P_{SHG}(2\omega)}{16\sqrt{2}S\omega^2P_{FF}^2(\omega)}}$$
Where $c$ is the speed of light, $\epsilon_0$ is the free space permittivity, $f=$\SI{76}{\mega\hertz} is the repetition rate of the pump laser, $r=$\SI{2.5}{\micro\meter} is the beam radius, $t_{FWHM}=$\SI{150}{\femto\second} is the FWHM of the pulse, $n_2=1.45$ is the refractive index of the quartz substrate, $P_{SHG}(2\omega)$ is the measured SH power, $S=0.94$ is the Gaussian shape factor and $P_{FF}^2(\omega)$ is the measured FB power. The calculated $\chi_S^{(2)}$ was found to be \SI{0.091}{\nano\metre\squared\per\volt}, is in good agreement with values reported in literature~\cite{rosa2018characterization,seyler2015electrical}. From this, we have calculated the generated SH power from the ATFL upon illumination with a FB beam diameter of \SI{100}{\micro\meter}, while taking into account that half of the $\ce{WSe2}$ monolayer area has been etched away. By comparing the calculated SH power with the detected power at the focal point, and considering all the losses of the setup, we retrieved the nonlinear focusing efficiency.

\subsection{Rayleigh-Sommerfeld Integral Formulation of the Zone Plate Lens Field}

According to scalar diffraction theory, the field at a point P$_0$ above an aperture $\Sigma$ can be expressed as function of the incident field on the aperture U(P$_1$) as:

\begin{equation}
U(P_0) = \frac{1}{j\lambda}
\iint_{\Sigma}
U(P_1)\,
\frac{\exp\left(j k r_{01}\right)}{r_{01}}\,
\cos\theta\,
\mathrm{d}s
\end{equation}

where $\lambda$ is the wavelength of SHG, $\overrightarrow{r_{01}}$ is the vector connecting P$_1$ to P$_0$ and $\theta$ is the angle between $\overrightarrow{r_{01}}$ and the normal to the $\Sigma$ plane\cite{GoodmanFourierOptics}. For calculation of the zone plate lens field, we take into account the radial symmetry of the lens, simplifying the integral:

\begin{equation}
U(P_0) = \frac{2\pi}{j\lambda}
\int_0^{\infty}
M(\rho)U(\rho)\,
\frac{\exp\left(j k r_{01}\right)}{r_{01}}\,
\cos\theta\,
\rho\mathrm{d}\rho
\end{equation}

where M($\rho$) is 0 in the rings and 1 outside of them, also setting the integration limit. For the illumination scheme presented in Fig. 3a, we illuminate the lens with a Gaussian beam of fundamental wavelength $\lambda_f$, so we write the pump field at the lens plane as:

\begin{equation}
U(\rho) = e^{-2\rho^2/w^2}e^{4\pi j \left(R_c-\sqrt{R_c^2-\rho^2}\right)/\lambda_f}
\end{equation}

where $w$ is the Gaussian beam waist and $R_c$ the phase radius of curvature, both set by the illumination objective.

\section{ACKNOWLEDGMENTS}
G.S. acknowledges funding from the Deutsche Forschungsgemeinschaft (DFG, German Research Foundation) - CRC/SFB 1375 NOA “Nonlinear Optics down to Atomic scales” (project number 398816777), International Research Training Group 2675 “META-Active” (project number 437527638 and project number 448835038), and WHAT-A-TWIST (project number 547611111). The van de Groep team acknowledges funding from the Open Technology Program of the Dutch National Science Foundation (NWO), grant number 19486, a Vidi grant (VI.Vidi.203.027) from the Netherlands Organization for Scientific Research (NWO), as well as an European Research Council Starting Grant under grant agreement No. 101116984. The authors acknowledge the use of OpenAI GPT 4.1 (accessed August 2026) to reorganize the different sections of the manuscript for clarity and flow. All AI-assisted text was reviewed and revised by the authors to ensure accuracy and clarity of meaning.

\section{Conflict of Interest}
The authors declare no conflicts of interest.

\section{Data Availability Statement}
The data that supports the findings of this study are available from the corresponding author upon reasonable request.

\section{AUTHOR CONTRIBUTION}
G.S. and R.R. conceived the work. R.R., B.D., T.H., M.A.D., B.V. and Z.L. fabricated and characterized the sample. R.R. and B.D. did the numerical simulations. R.R and D.M. designed and developed the optical setup. R.R. and T.W. performed the measurements on ATFL. R.R., B.D. and O.G. analyzed the data. G.S., J.G., Y.L. and C.R. supervised the project. R.R., B.D. and G.S. wrote the manuscript with contributions from all authors.

\clearpage

\clearpage

\clearpage

\clearpage

\clearpage

\bibliographystyle{naturemag}
\bibliography{Literature.bib}

\end{document}